\documentclass[trackchanges]{aastex7}

\usepackage{graphicx}	% Including figure files
\usepackage{amsmath}
\usepackage{amssymb}	% Extra maths symbols
\usepackage{longtable}
\usepackage{threeparttable}
\usepackage{multirow}
\usepackage{booktabs}
\usepackage{array}
\usepackage[figuresright]{rotating}
\usepackage[T1]{fontenc}   

\received{October 14, 2025}
\revised{October 26, 2025}
\accepted{November 14, 2025}

\begin{document}

\title{New Evidence for Extragalactic Einstein Probe Transients associated with Long Gamma-ray Bursts}

\author{Qin-Mei Li}
\affiliation{Department of Astronomy, School of Physics and Astronomy, Yunnan University, Kunming 650091, China}
\email[show]{qinmli@163.com (Q. M, Li)} 

\author[0000-0003-0516-404X]{Qi-Bin Sun}	
\email{sunqibin@ynu.edu.cn}
\affiliation{Department of Astronomy, School of Physics and Astronomy, Yunnan University, Kunming 650091, China}

\author{Sheng-Bang Qian}
\affiliation{Department of Astronomy, School of Physics and Astronomy, Yunnan University, Kunming 650091, China}
\email[show]{qiansb@ynu.edu.cn (S. B, Qian)}  

\author{Fu-Xing Li}
\affiliation{Department of Astronomy, School of Physics and Astronomy, Yunnan University, Kunming 650091, China}
\email{}

\begin{abstract}
The origin of extragalactic fast X-ray transients (EFXTs) remains a fundamental open question in high-energy astrophysics. The Einstein Probe (EP) mission provides a transformative opportunity to investigate their nature. While mounting observations of EP-discovered EFXTs (EP-EFXTs) suggest a possible connection to long gamma-ray bursts (lGRBs), an in-depth comparative analysis between them remains lacking. Here, we present a comparative analysis of their cosmic formation histories, revealing that EP-EFXTs and lGRBs share a similar evolutionary trend—showing a marked decline at $z < 1.0$ and a plateau beyond $1.0<z < 5$—which clearly distinguishes them from short GRBs. 
This result is derived from a rigorously selected sample of EP-EFXTs, using Lynden-Bell's $c^{-}$ method to reconstruct, for the first time, the luminosity function and formation rate of EP-EFXTs without any assumptions. Our findings provide independent evidence that EP-EFXTs and lGRBs may originate from a common progenitor channel.

\end{abstract}

\section{Introduction} \label{sec:intro}
Despite the discovery of extragalactic fast X-ray transients (EFXTs)—characterized by cosmological distances and short durations (seconds to hours)—in archival data from Chandra and XMM-Newton, the nature and origin of these sources continue to elude astronomers. The successful launch of the Einstein Probe (EP) in 2024 has opened up new avenues for investigating the origins of EFXTs.
A growing body of observational evidence suggests that the EFXTs discovered by the EP (EP-EFXTs) may possess afterglow properties and energetics similar to those of Gamma-Ray Bursts (GRBs). For instance, following the detection of EP240219a by EP \citep{2024ATel16463....1Z,2024ATel16472....1Z}, a coincident, faint, non-triggered gamma-ray transient was identified. Given their spatial and temporal consistency, \cite{2024ApJ...975L..27Y} and \citet{2024GCN.35784....1D} reclassified the event as GRB 240219A. Similarly, EP240315a was confirmed to be temporally and spatially coincident with GRB~240315C, establishing it as a unique GRB initially detected by a soft X-ray instrument \citep{2025NatAs...9..564L}. On the Amati relation, its position aligns with other high-redshift GRBs at the high-energy end, further reinforcing its consistency with the typical long GRBs (lGRBs) population. The study by \citet{2025ApJ...979L..28R} indicates that EP240315a was produced by a collimated jet with an opening angle of $\theta_{\rm j} \sim 3^\circ$ and a total energy of $E \sim 4 \times 10^{51}$~erg, properties consistent with those of lGRBs.
 Furthermore, \citet{2025NatAs...9.1073S} proposed that EP240414a originated from a weak relativistic jet associated with the Type Ic-BL supernova SN~2024gsa. This transient exhibited a radio luminosity comparable to that of bright GRBs, along with a broad spectral profile strikingly similar to other GRB-associated supernovae. \citet{2025A&A...701A.225B} found that the observed X-ray plateaus in EP240414a and EP241021a are consistent with an extension of the typical rest-frame correlations between GRB X-ray plateau luminosity and duration (Dainotti relation), supporting their interpretation within the GRB framework, and the observations of EP240801a can be explained by either an off-axis narrow jet or an intrinsically weak jet when placed on-axis \citep{2025ApJ...988L..34J}.

In a recent study, \citet{2025arXiv250907141O} curated a clean sample of approximately 113 EFXTs from EP/WXT detections. Based on this sample, they performed a comparative analysis of the cumulative redshift distribution between EP-EFXTs and lGRBs. The statistical tests indicate that their redshifts are drawn from same distributions. Furthermore, they found that the relation between spectral peak energy and isotropic-equivalent energy (Amati relation) is similar for both EP-EFXTs and lGRBs. They suggest that most EP-EFXTs are closely linked to lGRBs and originate from a massive star collapsar progenitor channel. 
However, the GRBs with different progenitors can occupy overlapping regions in the Amati and Dainotti relation planes \citep{2023MNRAS.524.1096L}. The redshift evolution of the luminosity of transients providing a key constrain for distinguishing between them (eg., \citealp{2024ApJ...973L..54C}). Therefore, building upon the clean sample from \citet{2025arXiv250907141O}, we conduct the first investigation into the formation rates (FR) of these two transient classes, providing new evidence for the connection between EP-EFXTs and lGRBs. 

This paper is structured as follows. In Section~2, we describe the EP-EFXTs sample and Lynden-Bell's $c^{-}$ method used in this study. The derived result for luminosity function (LF) and FR are presented in Section~3. Our conclusions and discussion are provided in Section~4. Throughout this work, we adopt a flat $\Lambda$CDM cosmology with $\Omega_{\mathrm{m}} = 0.3$ and $H_0 = 70~\mathrm{km~s^{-1}~Mpc^{-1}}$.

\section{Data and method}\label{sec:Results}
The EP launched in 2024, is a space mission in time-domain high-energy astrophysics led by the Chinese Academy of Sciences in collaboration with the European Space Agency and the Max Planck Institute for Extraterrestrial Physics in Germany \citep{2022hxga.book...86Y,2025SCPMA..6839501Y}. Its Wide-field X-ray Telescope (WXT), utilizing innovative lobster-eye micropore optics, achieves an exceptionally large instantaneous field of view of approximately 3,600 square degrees, with a detection sensitivity better than $\sim 3 \times 10^{-11} \ {\rm erg \ cm^{-2} \ s^{-1}}$ in the 0.5--4 keV band for a 1 ks exposure. The discoveries of these EP/WXT transients are typically distributed via Notices and Circulars of the General Coordinate Network (GCN). Designed to systematically discover and rapidly locate various X-ray transients, EP enables in-depth detection of high-energy transient phenomena such as faint and high-redshift GRBs, paving new avenues for understanding their emission mechanisms and progenitor nature.

\citet{2025arXiv250907141O} compiled a sample of EP-EFXTs with redshifts from public GCN Circulars \footnote{\url{https://gcn.nasa.gov/circulars}} and the published literature. Their selection criteria involved removing contaminants such as flare stars and Galactic transients, resulting in a clean sample of approximately 113 sources (for details, see \citealp{2025arXiv250907141O}). Among these, 26 sources had well-measured redshifts. For the subsequent analysis, the availability of a time-averaged X-ray flux was required, which led to the removal of three sources (EP241107a, EP250215a, and EP250226a). The final sample of 23 sources was used to compute the LF and FR. In Table \ref{tab:1}, we compile the available information of fast X-ray transient event, including the name of EP transient, the X-ray photon index $\Gamma_{WXT}$, time-averaged X-ray flux $F_{wxt,avg}$ in 0.5 -4 KeV band and X-ray luminosity $L_X$. 

\begin{figure}
	\centering
	\includegraphics[width=0.45\columnwidth]{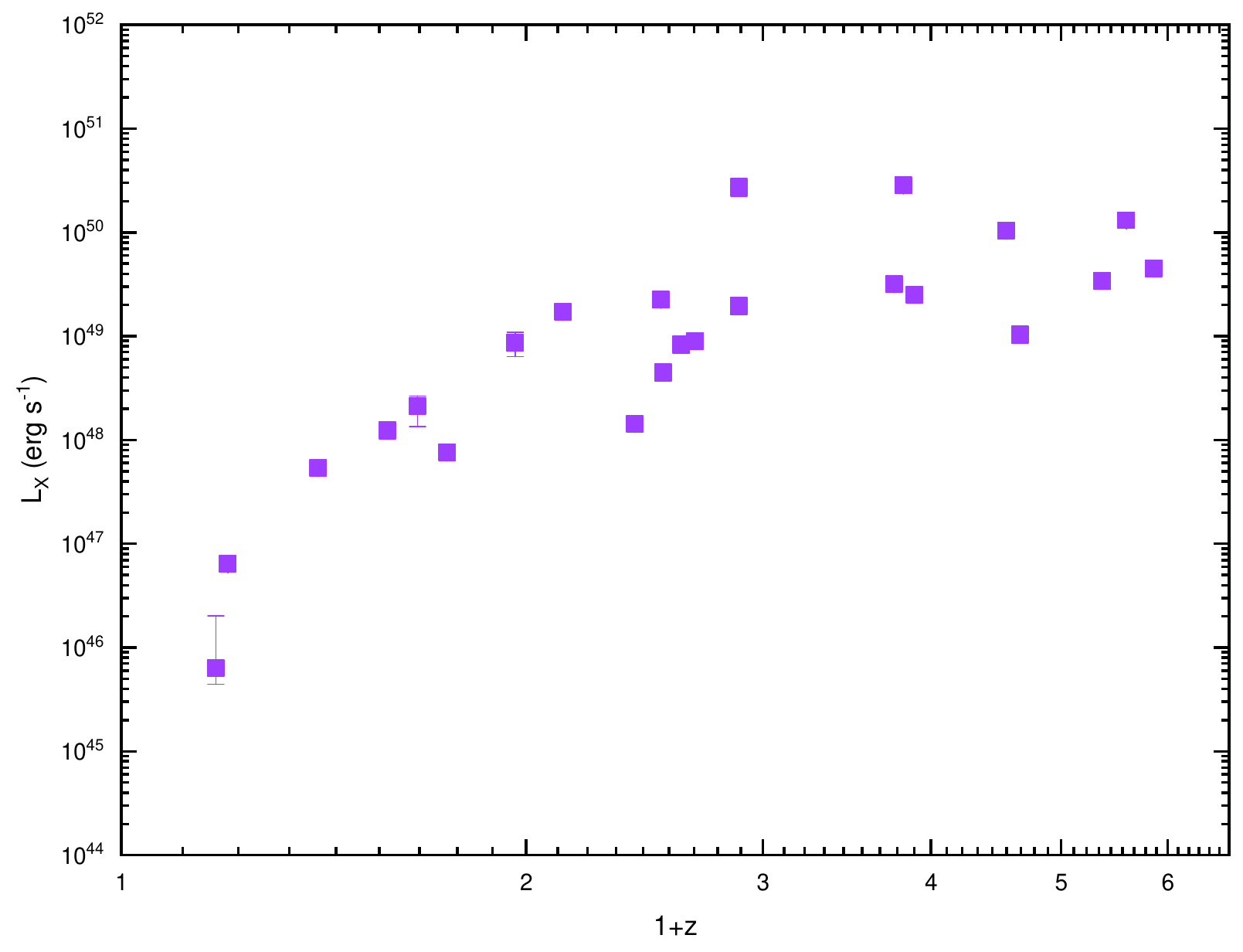}
	\includegraphics[width=0.45\columnwidth]{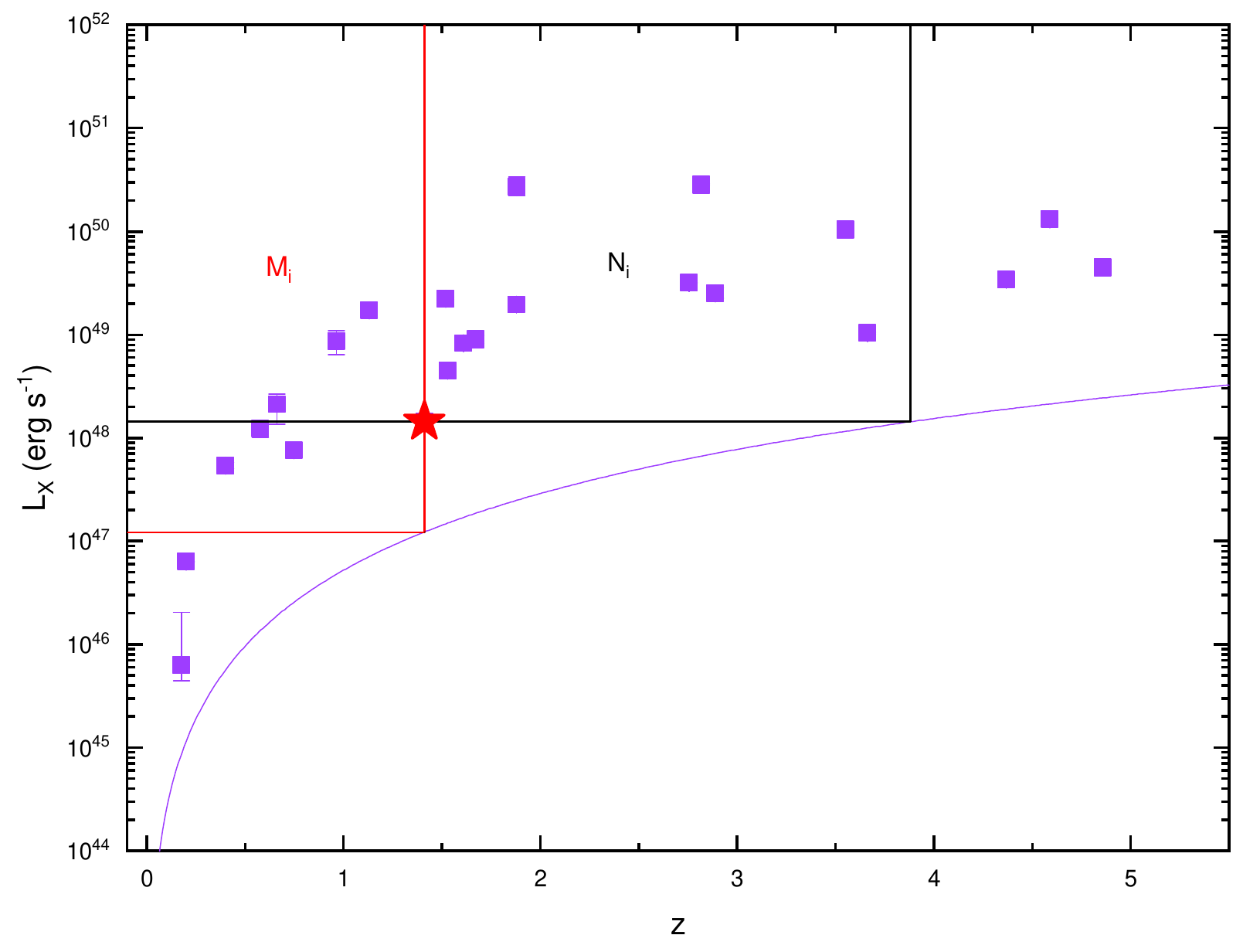}
	\includegraphics[width=0.45\columnwidth]{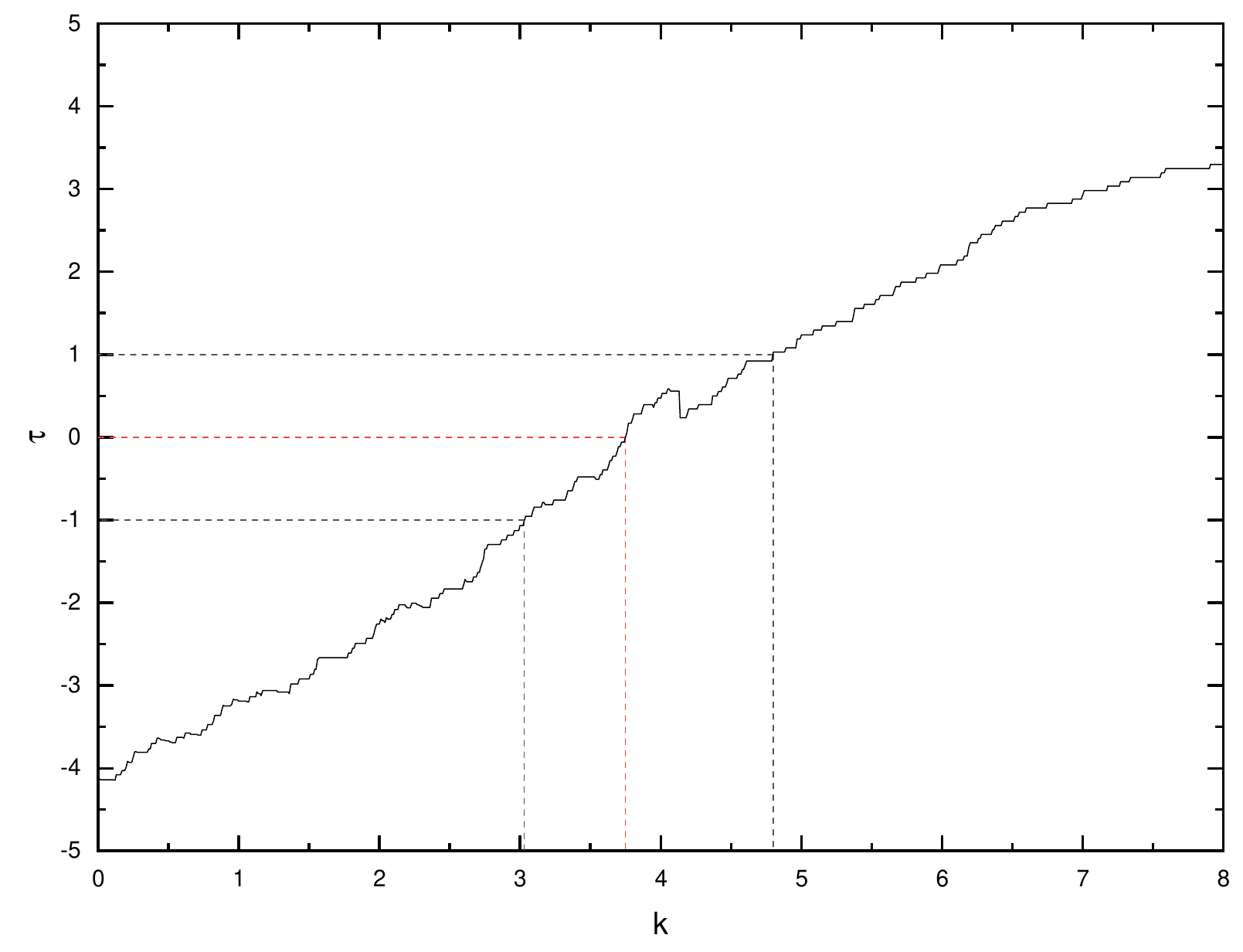}
	\includegraphics[width=0.45\columnwidth]{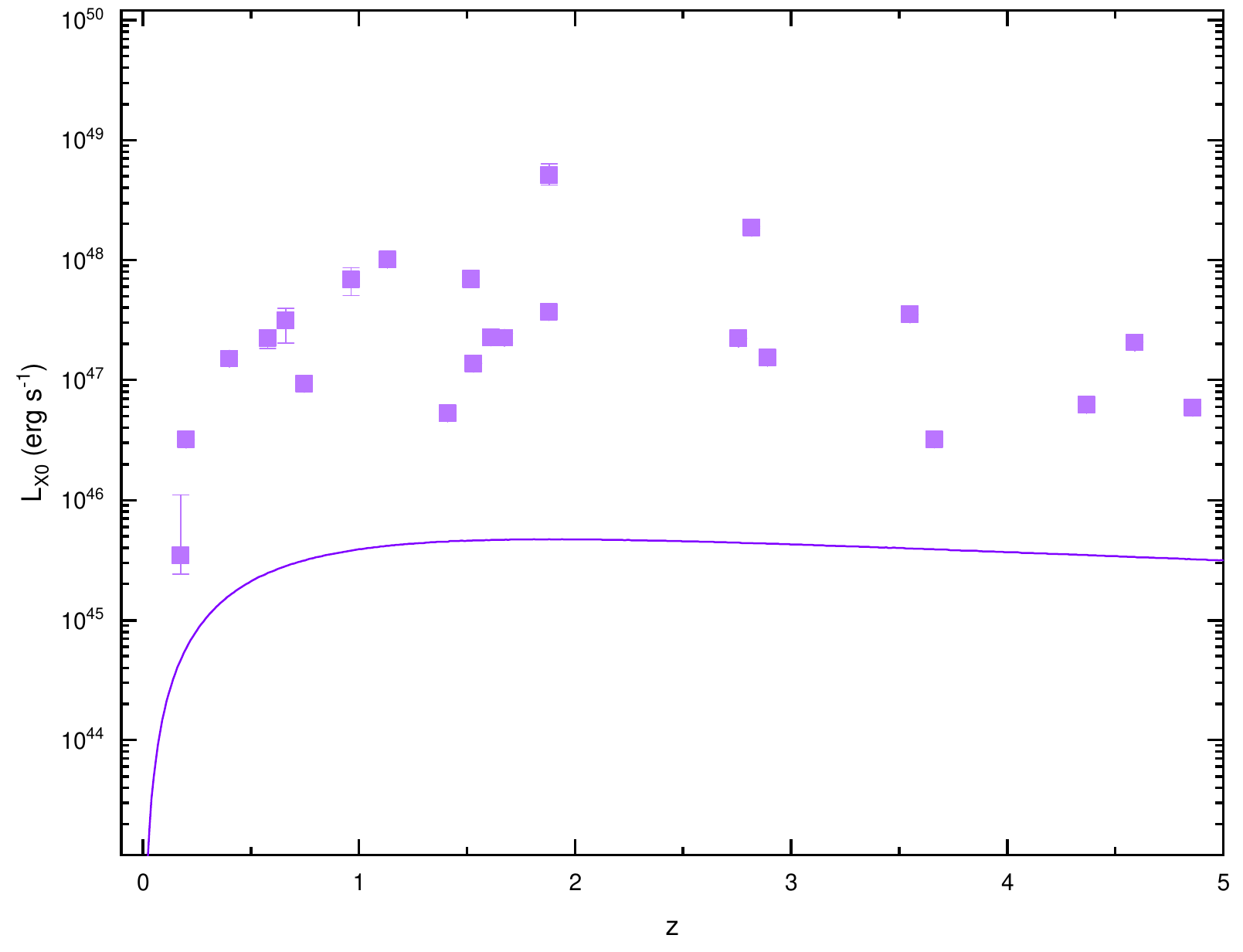}
	\caption{Distributions and correlations for the EP-EFXTs sample: 
		Left top: The redshift is correlation with X-ray luminosity; 
		Right top: X-ray luminosity distribution where individual points represent different EP-EFXTs, with the line indicating the sensitivity limit of $1.0 \times 10^{-11}\,\mathrm{erg\,cm^{-2}\,s^{-1}}$; 
		Left bottom: In the Kendall $\tau$ correlation test, the red dotted line represents the null hypothesis ($\tau = 0$), and the measured correlation strength of $k = 3.75$ suggests that the evolutionary dependence between luminosity and redshift has been effectively removed; 
		Right bottom: De-evolved LF following $L = L_{0}(1 + z)^{3.75}$ for our sample of 23 EP-EFXTs, removing the redshift evolution component.}
	\label{fig:1}
\end{figure}

The $L_X$ can be calculate by 
\begin{equation}
L_X = 4\pi d_L^2 F \cdot K
\end{equation}
where $d_L$ is the luminosity distance at redshift z for an assumed cosmological model, and the K is the K correction factor $K = (1 + z)^{\Gamma - 2}$. To ensure that all transients in the sample are above the flux limit, we adopt the lowest flux $F_{\rm limit} = 1 \times 10^{-11} \, \text{erg cm}^{-2} \text{s}^{-1}$ for EP-EFXTs as the flux threshold for the entire sample (see Figure \ref{fig:1} (b)). Hence, the corresponding luminosity limit at redshift $z$ can be calculated using ${L_{\rm limit}} = 4\pi d_L^2(z) F_{\rm limit}$.

The observed redshift distribution and inferred event rates of GRBs are subject to several selection effects \citep{2007NewAR..51..539C}, the most prominent of which stems from the satellite's instrumental limitations. Due to the existence of a detection flux threshold, GRBs below a certain brightness remain undetected. This leads to a truncated observed sample, making it impossible to reliably reconstruct the intrinsic redshift distribution without first correcting for such observational biases. Therefore, we compute the LF and event rate of EP-EFXTs directly from the observational data using Lynden-Bell's $c^{-}$ method.
\citet{1971MNRAS.155...95L} proposed a non-parametric approach devised to correct for flux-limit selection effects without a prior assumptions. Its applicability spans diverse astrophysical populations, including quasars \citep{2016Ap&SS.361..138D,2021ApJ...913..120Z}, GRBs \citep{2015ApJ...806...44P,2024MNRAS.527.7111L,2025ApJ...978..160L,2025ApJ...990L..54L}, and Fast radio bursts (FRBs) \citep{2024ApJ...973L..54C,2025ApJ...988L..64C}.

The $c^-$ method resolves the degeneracy between the LF and FR by explicitly extracting luminosity evolution. Ideally, if $L$ and $z$ in $\Psi(L, z)$ were independent, one could write $\Psi(L, z) = \psi(L)\phi(z)$ \citep{1992ApJ...399..345E}. Motivated by the known redshift-luminosity correlation in GRBs \citep{2004ApJ...609..935Y}, we searched for a similar relation in EP-EFXTs. The result shows a robust Spearman correlation coefficient of 0.86, revealing the presence of a parallel correlation in the EP-EFXT population (see Figure \ref{fig:1} (a), making the correct form $\Psi(L, z) = \psi_z(L)\phi(z)$. The solution is to remove this luminosity evolution $g(z)=(1+z)^k$ by defining a de-evolved luminosity $L_0 = L/g(z)$, making $L_0$ independent of $z$. This restores the factorizable form: $\Psi(L_0, z) = \phi(z)\psi(L_0)$. Transforming back via $L = L_0 g(z)$, we finally express the original distribution as $\Psi(L, z) = \psi_z(L)\phi(z) = \psi(L_0)\phi(z)$, in which the $\psi_z(L) = \psi(L / g(z))$.

The first step is obtain the form of the luminosity evolution $g(z)$. In Figure \ref{fig:1} (b), the $\textit{i}$th data in the $(L, z)$ plane, we can define \textit{$J_i$} as $J_i = \left\{ j \mid L_j \geq L_i,\ z_j \leq z_i^{\max} \right\}$, where $L_i$ denotes the luminosity of the $i$-th EP-EFXTs, and $z_i^{\max}$ is the maximum redshift at which a EP-EFXTs of luminosity $L_i$ can be detected. This region corresponds to the black rectangle in Figure~\ref{fig:1} (b). Let $n_i$ be the number of EP-EFXTs contained in $J_i$. Then $N_i = n_i - 1$ (i.e., excluding the $i$-th EP-EFXTs itself) is equivalent to the quantity $c^{-}$ introduced by \citet{1971MNRAS.155...95L}.
Similarly, we define the set $J_i'$ as $J_i' = \left\{ j \mid L_j \geq L_i^{\lim},\ z_j < z_i \right\},$
where $L_i^{\lim}$ is the minimum luminosity detectable at redshift $z_i$. This region is indicated by the red rectangle in Figure~\ref{fig:1} (b), and contains $M_i$ events.

Now consider the $n_i$ EP-EFXTs in the black rectangle of Figure \ref{fig:1} (b). Let $R_i$ be the number of events with redshift less than or equal to $z_i$. Under the assumption that luminosity $L$ and redshift $z$ are independent, $R_i$ follows a uniform distribution over the integers $1$ to $n_i$ \citep{1992ApJ...399..345E}. The test statistic $\tau$ is then defined as
$\tau \equiv \sum_i \frac{R_i - E_i}{\sqrt{V_i}}$ with $E_i = (1 + n_i)/2$ and $V_i = (n_i^2 - 1)/12$ representing the expectation and variance of $R_i$, respectively. If $R_i$ is indeed uniformly distributed between $1$ and $n_i$, the counts of $R_i \leq E_i$ and $R_i \geq E_i$ should be nearly equal, yielding $\tau \approx 0$. Therefore, by choosing a luminosity evolution function $g(z)$ such that $\tau = 0$, we can remove the effect of luminosity evolution via the transformation $L_0 = \frac{L}{g(z)}$.

To determine the value of $k$ that ensures the independence between the de-evolved luminosity $L_0 = L / g(z)$ and redshift $z$, we vary $k$ until the test statistic $\tau$ approaches zero. As shown in Figure~\ref{fig:1} (c), the best-fit value is $k = 3.75^{+1.05}_{-0.72}$ at the $1\sigma$ confidence level. Consequently, the luminosity evolution function is taken as $g(z) = (1 + z)^{3.75}$.

After converting the observed luminosity to the de-evolved luminosity $L_0 = L / (1 + z)^{3.75}$, we can derive the local cumulative LF $\psi(L_0)$ using the non-parametric method based on the following equation \citep{1971MNRAS.155...95L,1992ApJ...399..345E}:

\begin{equation}
	\psi(L_{0i}) = \prod_{j < i} \left( 1 + \frac{1}{N_j} \right),
\end{equation}

\noindent where $j < i$ denotes that the $j$-th EP-EFXTs has a de-evolved luminosity $L_{0j}$ greater than $L_{0i}$. Similarly, the cumulative redshift distribution $\phi(z)$ is given by

\begin{equation}
	\phi(z_i) = \prod_{j < i} \left( 1 + \frac{1}{M_j} \right),
	\label{equ:1}
\end{equation}

\noindent where $j < i$ now indicates that the $j$-th EP-EFXTs has a redshift $z_j$ smaller than $z_i$.

\section{Results}\label{sec:Results}
\begin{figure}
	\centering
	\includegraphics[width=0.45\columnwidth]{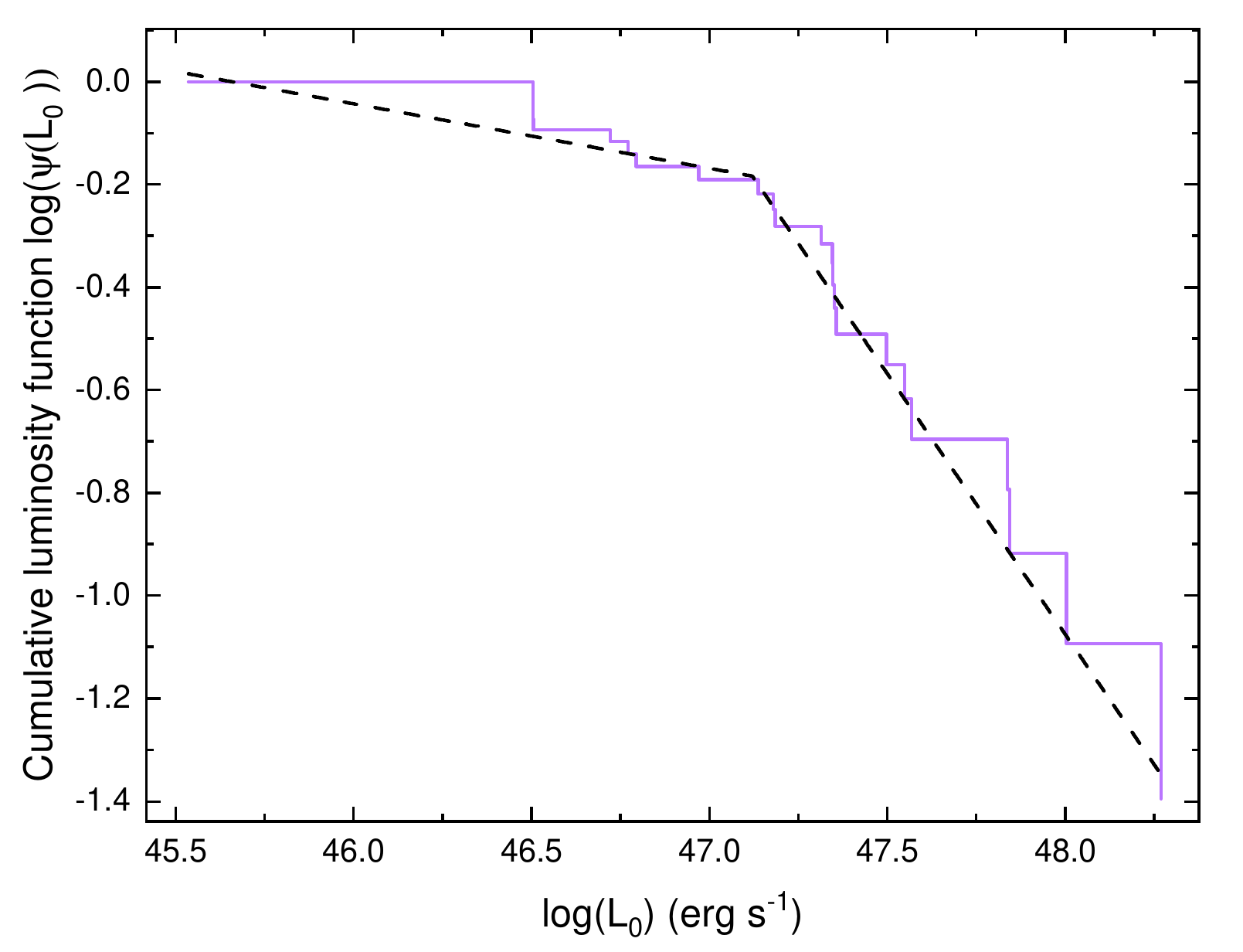}
	\includegraphics[width=0.45\columnwidth]{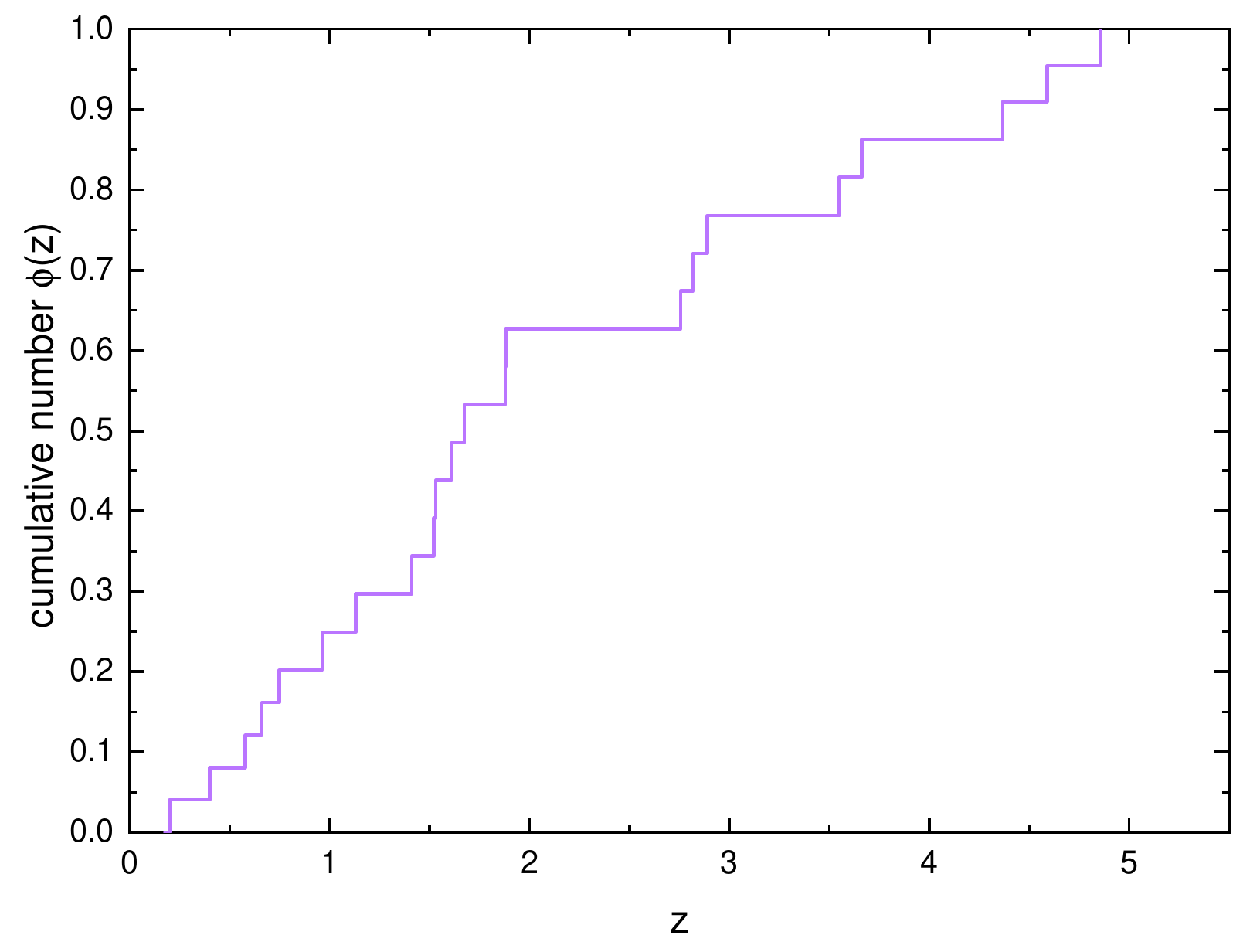}
	\caption{Left: The distribution of cumulative LF of EP-EFXTs; Right: Normalized cumulative redshift distribution.
		.\label{fig:2}}
\end{figure}

As described in the previous section, the luminosity evolution function is determined as $g(z) = (1+z)^{3.75}$ via the non-parametric $\tau$ test method. The de-evolved luminosity is defined as $L_0 = L / g(z)$, whose distribution is presented in Figure~\ref{fig:1}. Using this updated dataset, we derive the local cumulative LF $\psi(L_0)$ via Lynden-Bell's $c^{-}$ method, as shown in Figure~\ref{fig:2} (a). The resulting LF $\psi(L_0)$ after removing redshift evolution is well described by a broken power-law form:
\begin{equation}
	\psi(L_0) \propto 
	\begin{cases}
		L_0^{\,-0.13 \pm 0.04}, & L_0 < L_0^b, \\
		L_0^{\,-1.01 \pm 0.04}, & L_0 > L_0^b,
	\end{cases}
\end{equation}
where $L_0^b = 1.31 \times 10^{47}~\text{erg~s}^{-1}$ is the break luminosity. We note that $\psi(L_0)$ represents the local LF at $z = 0$, as luminosity evolution has been removed. The LF at an arbitrary redshift $z$ can be recovered via
$\psi_z(L) = \psi\left( L / g(z) \right) = \psi\left( L / (1+z)^{3.75} \right)$.
Accordingly, the break luminosity at redshift $z$ evolves as $L_z^b = L_0^b \cdot (1+z)^{3.75}$.

We are often more interested in the comoving differential form of $\phi(z)$, which corresponds to the cosmic FR of EP-EFXTs, $\rho(z)$. This rate can be expressed as

\begin{equation}
	\rho(z) = \frac{d\phi(z)}{dz} \cdot (1+z) \cdot \left( \frac{dV(z)}{dz} \right)^{-1},
	\label{equ:fr}
\end{equation}

\noindent where the factor $(1+z)$ accounts for cosmological time dilation, and $dV(z)/dz$ is the differential comoving volume. The latter is given by

\begin{equation}
	\frac{dV(z)}{dz} = 4\pi \left( \frac{c}{H_0} \right)^3
	\left( \int_0^z \frac{dz'}{\sqrt{ \Omega_\Lambda + \Omega_m (1+z')^3 }} \right)^2
	\cdot \frac{1}{\sqrt{ \Omega_\Lambda + \Omega_m (1+z')^3 }}.
\end{equation}

\begin{figure}
	\centering
	\includegraphics[width=0.9\columnwidth]{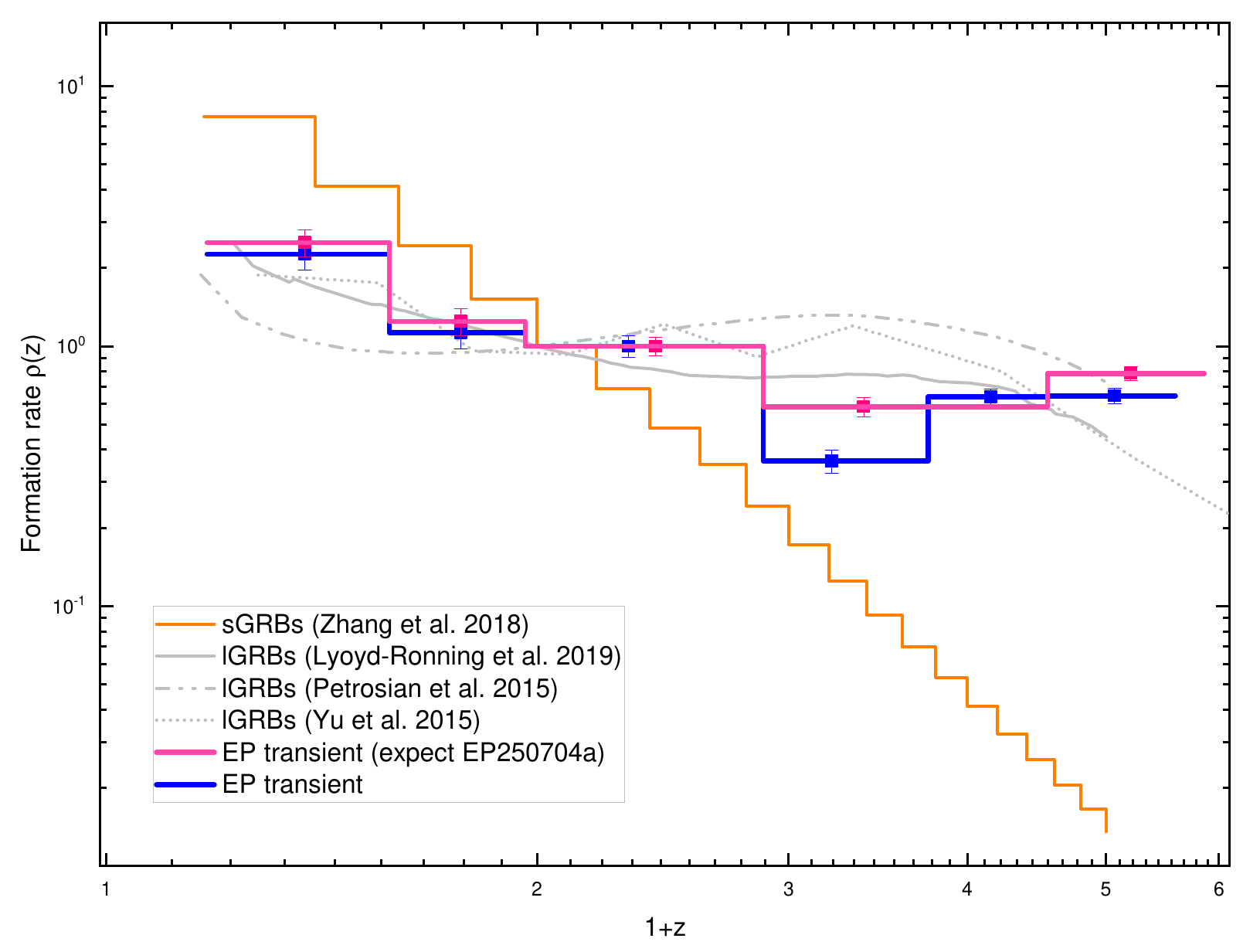}
	\caption{Comparison rate between EP-EFXTs and lGRBs, sGRBs. The lGRB rate and sGRB rate were collected from \citep{2015ApJ...806...44P,2015ApJS..218...13Y,2019MNRAS.488.5823L} and \citet{2018ApJ...852....1Z}. The blue line are the FR of 23 EP-EFXTs (pink line exclude the EP250704a). The error bar gives a 1 $\sigma$ error. These fitting lines are normalized at z=1.}
	\label{fig:3}
\end{figure}

The cumulative redshift distribute is presented in Figure~\ref{fig:2} (b). To convert this into a cosmic FR, the differential term $d\phi(z)/dz$ must be evaluated, as specified in Equation~\ref{equ:fr}. The comoving FR $\rho(z)$, derived from Equation~\ref{equ:fr}, is depicted in Figure~\ref{fig:3}. Here, the blue histogram shows the redshift evolution of the EP-EFXTs rate, with $1\sigma$ confidence limits. The FR decreases sharply at $z < 1$, remains nearly flat for $1<z<5$. We compared our results with earlier studies on the FR of lGRBs \citep{2015ApJ...806...44P,2015ApJS..218...13Y,2019MNRAS.488.5823L} and sGRBs \citep{2018ApJ...852....1Z}. We found that the EP transient rate exhibit distinct evolutionary trends when compared to long and short GRB rates. The lGRBs rate is broadly consistent with that of EP-EFXTs,  whereas the sGRB rate diverges significantly, pointing to massive core collapse as their common progenitor.

%%%%%%%%%%%%%%%%%%%%%%%%%%%%%%%%%%%%DISCUSSION%%%%%%%%%%%%%%%%%%%%%%%%%%%%%%%%%%%%%%%%
\section{Conclusions and Discussions} \label{sec:DISCUSSION}
In this study, we apply Lynden-Bell's $c^{-}$ method to investigate the LF and FR of EP-EFXTs without any assumptions. We first employ the $\tau$ statistical method to remove luminosity evolution by adopting an evolutionary form $g(z) = (1+z)^k$. The evolves value is found to be $k = 3.75^{+1.05}_{-0.72}$, which yields $\tau = 0$. After correcting for luminosity evolution via $L_0 = L / (1+z)^{3.75}$. We further derive the EP-EFXTs FR from the differential form of $\phi(z)$. Using Equation~\ref{equ:fr}, we obtain the cosmic GRB FR $\rho(z)$, shown in Figure~\ref{fig:3}. Our results indicate a decline in the EP-EFXT FR at $z < 1.0$, while it remains constant at higher redshifts $1<z <5$. The EP-EFXT FR $\rho(z)$ agrees with the observed FR of lGRBs \citep{2015ApJ...806...44P,2015ApJS..218...13Y,2019MNRAS.488.5823L} and found that they exhibit broadly consistent evolutionary trends. At the same time, we also compared the event rates of EP-EFXTs and sGRBs \citep{2018ApJ...852....1Z} and found that their trends are significantly different. Our results and conclusions are not affected by contamination from compact binary mergers. Among our 23 samples, 10 of them are confirmed to correspond with GRBs, of which 9 are associated with lGRBs and 1 with sGRB \citep{2025GCN.40972....1F}. After removing this EP250704a event associated with the sGRB, we recalculated the event rate of EP-EFXTs and found almost no change, shown as in Figure \ref{fig:3}.

With the launch of the EP, the study of the origin of such EFXTs has entered a new phase. A growing body of observational evidence indicates that the EP-EFXTs exhibit multiple key properties consistent with GRBs, including spatial–temporal coincidence (e.g., EP240219a and EP240315a; \citealp{2024ATel16463....1Z,2024ATel16472....1Z,2024GCN.35784....1D,2025NatAs...9..564L}), conformity with the Amati relation, similar energetics and jet opening angles (\citealp{2025ApJ...979L..28R}), comparable radio luminosity (\citealp{2025NatAs...9.1073S}), and adherence to the Dainotti relation (as seen in EP240414a and EP241021a; \citealp{2025A&A...701A.225B}), redshift distrbution \citep{2025arXiv250907141O}. Furthermore, the spectral profiles of EP240414a resemble those of Type Ic supernovae (\citet{2025NatAs...9.1073S}), suggesting that GRBs and these EP-EFXTs may share a common progenitor.

The FR of extragalactic transients serves as a key diagnostic for understanding their progenitors and can be used to constrain their formation pathways (eg., the FR of fast radio bursts is similar to that of sGRBs, providing evidence that their progenitors may be compact binary mergers; \citealp{2024ApJ...973L..54C}). EP-EFXTs have been detected out to a redshift of approximately $z \sim 5$, a distance range similarly probed by lGRBs, which are known to extend to even higher redshifts. Furthermore, previous studies have suggested a potential physical connection between EP-EFXTs and lGRBs. Given these observational and theoretical contexts, it is reasonable to compare the FR of EP transients and lGRBs, sGRBs. This result provides new evidence for a connection between EP-EFXTs and lGRBs, suggesting that they may share a common progenitor. 

Given the nascent stage of the EP mission (less than two years in operation), our conclusions are necessarily based on a limited dataset and must be tested against the larger samples of EP-EFXTs anticipated in the coming years. Therefore, further research into the EP-EFXT formation rate with substantially increased statistics constitutes a particularly promising path forward.  
It is also noteworthy that although the original sample compiled by \citet{2025arXiv250907141O} contains 113 sources, the current analysis depends on redshift and flux measurements, limiting the usable sample to only 23 sources, or about 20\% of the original dataset. The remaining 80\% of the sample could introduce additional uncertainties. The redshift incompleteness may result from difficulties in performing rapid follow-up observations \citep{2025arXiv250907141O}, or from the intrinsically faint afterglows of sGRBs, which are challenging to detect without timely follow-up \citep{2014ARA&A..52...43B}. Therefore, the findings presented in this work necessitate validation using a larger sample with more complete redshift information.

\section*{Acknowledgements}
We extend our gratitude to the entire Einstein Probe mission team for their efforts. We are particularly indebted to O'Connor et al. for their statistical work on extragalactic X-ray transients, which made this project feasible. This work was supported by the Scientific Research Foundation of the Education Department of Yunnan Province, the Graduate Research Innovation Foundation of Yunnan University (KC-252511615), the Postdoctoral Fellowship Program of CPSF under Grant Number GZC20252095, the China Postdoctoral Science Foundation under Grant Number 2025M773194, Caiyun Postdoctoral Program in Yunnan Province of China (grant No. C615300504124), the National Natural Science Foundation of China (grant Nos. 12503040, 11933008 and 12303040), National Key R\&D Program of China (grant No. 2022YFE0116800), Yunnan Fundamental Research Projects (grant NOs. 202501AS070055, 202503AP140013, 202201AT070092 and 202401AT070143).

\bibliography{sample7}{}
\bibliographystyle{aasjournalv7}
\newpage
\renewcommand\thefigure{A1}

\setlength{\tabcolsep}{1mm}{
	\renewcommand\arraystretch{1}
	\begin{center}
		\begin{longtable}{lccccc}
			\caption{23 EP transient Included in Our analysis}
			\label{tab:1} \\
			\hline%
			EP &z  & $\Gamma$ & $F_{WXT}$  &$L_X$  & K    \\
			&   (s)         &     & $10 ^{-10}erg/cm ^{2}$/s &$erg\, s^{-1}$  &    \\
			(1)&(2)& (3) &(4) & (5) &(6) \\
			\hline%
			\endhead%
			\hline%
			\endfoot%
			\hline%
			\endlastfoot%
EP240315a&4.859 &1.40 &$5.30 ^{+1.00 }_{-0.70 }$&$4.474 ^{+0.071 }_{-0.050 }\times 10 ^{49}$&2.89  \\
EP240414a&0.401 &3.10 &$6.50 ^{+1.30 }_{-1.00 }$&$5.356 ^{+0.377 }_{-0.290 }\times 10 ^{47}$&0.69  \\
EP240801a&1.673 &2.00 &$4.80 ^{+3.10 }_{-3.10 }$&$8.959 ^{+0.810 }_{-0.810 }\times 10 ^{48}$&1.00  \\
EP240804a&3.662 &0.70 &$6.10 ^{+0.30 }_{-0.20 }$&$1.029 ^{+0.017 }_{-0.011 }\times 10 ^{49}$&7.40  \\
EP240806a&2.818 &2.60 &$19.00 ^{+18.00 }_{-6.00 }$&$2.827 ^{+0.082 }_{-0.027 }\times 10 ^{50}$&0.45  \\
EP241021a&0.748 &1.80 &$3.31 ^{+0.13 }_{-0.09 }$&$7.571 ^{+0.109 }_{-0.075 }\times 10 ^{47}$&1.12  \\
EP241030a&1.411 &2.50 &$0.75 ^{+0.03 }_{-0.03 }$&$1.428 ^{+0.006 }_{-0.006 }\times 10 ^{48}$&0.64  \\
EP241113a&1.530 &1.30 &$5.70 ^{+0.13 }_{-0.08 }$&$4.457 ^{+0.030 }_{-0.019 }\times 10 ^{48}$&1.92  \\
EP241217a&4.590 &1.90 &$7.30 ^{+0.30 }_{-0.30 }$&$1.311 ^{+0.002 }_{-0.002 }\times 10 ^{50}$&1.19  \\
EP241217b&1.879 &1.60 &$12.00 ^{+1.00 }_{-1.00 }$&$1.952 ^{+0.030 }_{-0.030 }\times 10 ^{49}$&1.53  \\
EP250108a&0.176 &2.80 &$0.64 ^{+2.25 }_{-0.30 }$&$6.311 ^{+14.092 }_{-1.879 }\times 10 ^{45}$&0.88  \\
EP250125a&2.890 &0.80 &$18.00 ^{+7.00 }_{-5.00 }$&$2.496 ^{+0.327 }_{-0.234 }\times 10 ^{49}$&5.10  \\
EP250205a&3.550 &2.50 &$4.20 ^{+0.11 }_{-0.11 }$&$1.039 ^{+0.001 }_{-0.001 }\times 10 ^{50}$&0.47  \\
EP250223a&2.756 &2.10 &$4.40 ^{+0.14 }_{-0.11 }$&$3.170 ^{+0.006 }_{-0.005 }\times 10 ^{49}$&0.88  \\
EP250302a&1.131 &0.60 &$70.00 ^{+20.00 }_{-16.00 }$&$1.722 ^{+0.312 }_{-0.250 }\times 10 ^{49}$&2.88  \\
EP250304a&0.200 &2.20 &$5.30 ^{+0.04 }_{-0.04 }$&$6.326 ^{+0.032 }_{-0.032 }\times 10 ^{46}$&0.96  \\
EP250321a&4.368 &0.66 &$17.00 ^{+2.00 }_{-2.00 }$&$3.393 ^{+0.132 }_{-0.132 }\times 10 ^{49}$&9.50  \\
EP250404a&1.88&0.40 &$590.00 ^{+220.00 }_{-160.00 }$&$2.700 ^{+0.659 }_{-0.480 }\times 10 ^{50}$&7.86  \\
EP250416a&0.963 &0.30 &$57.00 ^{+18.00 }_{-18.00 }$&$8.631 ^{+2.226 }_{-2.226 }\times 10 ^{48}$&3.15  \\
EP250427a&1.520 &1.70 &$20.00 ^{+3.00 }_{-2.00 }$&$2.233 ^{+0.070 }_{-0.046 }\times 10 ^{49}$&1.32  \\
EP250704a&0.661 &1.70 &$13.00 ^{+8.00 }_{-11.00 }$&$2.110 ^{+0.548 }_{-0.754 }\times 10 ^{48}$&1.16  \\
EP250821a&0.577 &1.20 &$13.00 ^{+4.00 }_{-4.00 }$&$1.226 ^{+0.218 }_{-0.218 }\times 10 ^{48}$&1.44  \\
EP250827a&1.610 &0.70 &$17.00 ^{+4.00 }_{-4.00 }$&$8.294 ^{+0.997 }_{-0.997 }\times 10 ^{48}$&3.48  \\
	
	\end{longtable}
	\end{center}

\end{document}